\documentclass[aps,pra,twocolumn,superscriptaddress]{revtex4}

\usepackage[normalem]{ulem}

\usepackage{mathpazo}

\usepackage{amsmath}
\usepackage{amsfonts, amssymb, amsthm, bbm, braket, accents}
\usepackage{graphicx}   
\usepackage{subfigure}  
\usepackage{amsbsy} 
\usepackage{color}
\usepackage[bold]{hhtensor} 
\usepackage{natbib}
\usepackage{algpseudocode}

\usepackage{multirow}

\usepackage{ifthen}
\newboolean{FinalVersion}
\setboolean{FinalVersion}{false}
\newboolean{JournalVersion}
\setboolean{JournalVersion}{false}

\ifthenelse{\boolean{JournalVersion}}{
  
}{
    \usepackage{mathpazo}
}

\ifthenelse{\boolean{FinalVersion}}{

}{

}

\newcommand{\contentfolder}{.}

\newcommand{\T}{\mathrm{T}}
\newcommand\Tr{\mathrm{Tr}}
\newcommand\supp{\operatorname{supp}}

\newcommand{\expect}{\mathbb{E}}
\newcommand{\ident}{\mathbbm{1}}
\newcommand{\ave}{\mathrm{ave}}
\newcommand{\rref}{\mathrm{ref}}

\newcommand{\ii}{\mathrm{i}}

\newcommand{\defeq}{:=}

\newlength{\dhatheight}
\newcommand{\hhatDS}[1]{%
    \settoheight{\dhatheight}{\ensuremath{\hat{#1}}}%
    \addtolength{\dhatheight}{-0.25ex}%
    \hat{\vphantom{\rule{1pt}{\dhatheight}}%
    \smash{\hat{#1}}}}
\newcommand{\hhatTS}[1]{%
    \settoheight{\dhatheight}{\ensuremath{\hat{#1}}}%
    \addtolength{\dhatheight}{-0.25ex}%
    \hat{\vphantom{\rule{1pt}{\dhatheight}}%
    \smash{\hat{#1}}}}    
\newcommand{\hhatS}[1]{%
    \settoheight{\dhatheight}{\ensuremath{\scriptstyle{\hat{#1}}}}%
    \addtolength{\dhatheight}{-0.175ex}%
    \hat{\vphantom{\rule{1pt}{\dhatheight}}%
    \smash{\hat{#1}}}}
\newcommand{\hhatSS}[1]{%
    \settoheight{\dhatheight}{\ensuremath{\scriptscriptstyle{\hat{#1}}}}%
    \addtolength{\dhatheight}{-0.07ex}%
    \hat{\vphantom{\rule{1pt}{\dhatheight}}%
    \smash{\hat{#1}}}}
\newcommand{\hhat}[1]{\mathchoice{\hhatDS{#1}}{\hhatTS{#1}}{\hhatS{#1}}{\hhatSS{#1}}}

\makeatletter
\def\Decl@Mn@Delim#1#2#3#4{%
  \if\relax\noexpand#1%
    \let#1\undefined
  \fi
  \DeclareMathDelimiter{#1}{#2}{#3}{#4}{#3}{#4}}
\def\Decl@Mn@Open#1#2#3{\Decl@Mn@Delim{#1}{\mathopen}{#2}{#3}}
\def\Decl@Mn@Close#1#2#3{\Decl@Mn@Delim{#1}{\mathclose}{#2}{#3}}
\DeclareFontFamily{OMX}{MnSymbolE}{}
\DeclareFontShape{OMX}{MnSymbolE}{m}{n}{
    <-6>  MnSymbolE5
   <6-7>  MnSymbolE6
   <7-8>  MnSymbolE7
   <8-9>  MnSymbolE8
   <9-10> MnSymbolE9
  <10-12> MnSymbolE10
  <12->   MnSymbolE12}{}
\DeclareFontShape{OMX}{MnSymbolE}{b}{n}{
    <-6>  MnSymbolE-Bold5
   <6-7>  MnSymbolE-Bold6
   <7-8>  MnSymbolE-Bold7
   <8-9>  MnSymbolE-Bold8
   <9-10> MnSymbolE-Bold9
  <10-12> MnSymbolE-Bold10
  <12->   MnSymbolE-Bold12}{}
\DeclareSymbolFont{mnsymbols}  {OMX}{MnSymbolE}{m}{n}
\Decl@Mn@Open {\lsem}               {mnsymbols}{'102}
\Decl@Mn@Close{\rsem}               {mnsymbols}{'107}
\Decl@Mn@Open {\llangle}            {mnsymbols}{'164}
\Decl@Mn@Close{\rrangle}            {mnsymbols}{'171}
\makeatother

\newcommand{\eq}[1]{\hyperref[eq:#1]{(\ref*{eq:#1})}}
\renewcommand{\sec}[1]{\hyperref[sec:#1]{Section~\ref*{sec:#1}}}
\newcommand{\app}[1]{\hyperref[app:#1]{Appendix~\ref*{app:#1}}}
\newcommand{\tab}[1]{\hyperref[tab:#1]{Table~\ref*{tab:#1}}}
\newcommand{\alg}[1]{\hyperref[alg:#1]{Algorithm~\ref*{alg:#1}}}
\newcommand{\fig}[1]{\hyperref[fig:#1]{Figure~\ref*{fig:#1}}}
\newcommand{\thm}[1]{\hyperref[thm:#1]{Theorem~\ref*{thm:#1}}}
\newcommand{\lem}[1]{\hyperref[lem:#1]{Lemma~\ref*{lem:#1}}}
\newcommand{\cor}[1]{\hyperref[cor:#1]{Corollary~\ref*{cor:#1}}}
\newcommand{\defn}[1]{\hyperref[def:#1]{Definition~\ref*{def:#1}}}

\usepackage[colorlinks=true]{hyperref} 
\usepackage{algorithm} 

\begin{document}


\title{Accelerated Randomized Benchmarking}

\author{Christopher Granade}
\email[]{cgranade@cgranade.com}
\homepage[]{http://cgranade.com/}
\thanks{\\
     Literate source code for this work is available at \url{https://github.com/cgranade/accelerated-randomized-benchmarking}. To view the code online, visit \url{http://nbviewer.ipython.org/github/cgranade/accelerated-randomized-benchmarking/blob/master/src/model_testing.ipynb}.
}
\affiliation{
    Institute for Quantum Computing,
    University of Waterloo,
    Waterloo, ON
}
\affiliation{
    Department of Physics,
    University of Waterloo,
    Waterloo, ON
}

\author{Christopher Ferrie}
\affiliation{
Center for Quantum Information and Control,
University of New Mexico,
Albuquerque, New Mexico, 87131-0001}

\author{D. G. Cory}
\affiliation{
    Institute for Quantum Computing,
    University of Waterloo,
    Waterloo, ON
}
\affiliation{
    Department of Chemistry,
    University of Waterloo,
    Waterloo, ON
}
\affiliation{
    Perimeter Institute,
    Waterloo, ON
}
\affiliation{
    Canadian Institute for Advanced Research,
    Toronto, Ontario
}

\date{\today}



\begin{abstract}
Quantum information processing offers promising advances for a wide range of
fields and applications, provided that we can efficiently assess the performance
of the control applied in candidate systems.
That is, we must be able to determine whether we have implemented a
desired gate, and refine accordingly.
Randomized benchmarking reduces the difficulty of this task by exploiting
symmetries in quantum operations.

Here, we bound the resources required for benchmarking and show that,
with prior information, we can achieve several orders of magnitude
better accuracy than in traditional approaches to benchmarking. Moreover,
by building on state-of-the-art classical algorithms, we
reach these accuracies with near-optimal resources. Our approach requires
an order of magnitude less data to achieve the same accuracies and
to provide online estimates of the errors in the reported fidelities.
We also show that our approach is useful for physical devices by comparing
to simulations.

Our results thus enable the application of randomized benchmarking in new
regimes, and dramatically reduce the experimental effort required to assess
control fidelities in quantum systems. Finally, our work is based on
open-source scientific libraries, and can readily be applied in systems of
interest.
\end{abstract}


\maketitle

Quantum information processing devices offer great promise in a variety of
different fields, including chemistry and material science, data analysis and machine learning~
\cite{wiebe_quantum_2012,wiebe_quantum_2014-1,kassal_simulating_2011,hastings_improving_2014},
as well as cryptography \cite{shor_polynomial-time_1995}.
Over the past few years, proposals have been advanced for quantum information
processing past the classical scale, based on node-based architectures
\cite{nickerson_freely_2014,borneman_parallel_2012}.
In addition, rapid progress has been made towards
experimental implementations that might
allow for developing such devices \cite{chow_implementing_2014,barends_logic_2014}.
An impediment in this effort, however,
is presented by the difficulty of calibrating and diagnosing quantum devices.

In particular, in the development of
quantum information processing, an important experimental challenge
is to efficiently characterize the quality with which we can control
a quantum system.
By characterizing the quality of a quantum gate that is implemented by
a control pulse, we can then reason
about the utility of that gate for quantum information processing
tasks.
For instance, we can estimate the feasibility of and the resources required
to implement error correction using that control by comparing to
proven and numerically estimated fault-tolerance thresholds
\cite{gottesman_introduction_2009,fowler_high-threshold_2009}, or can adjust our
control sequences to account for differences between our control model and
the actual system.

In cases where only the quality of a quantum gate or set of gates is required, randomized
benchmarking has proven to be a useful means of extracting
this information with relatively little
experimental effort
\cite{magesan_characterizing_2012}, has been demonstrated in a variety
of experimental settings
\cite{ryan_randomized_2009,chow_randomized_2009,olmschenk_randomized_2010,
brown_single-qubit-gate_2011,
moussa_practical_2012,
tan_demonstration_2013,
corcoles_process_2013,barends_logic_2014}.
Randomized benchmarking has also been used
to improve gate fidelities by
characterizing cross-talk
\cite{gambetta_characterization_2012} or distortions
\cite{gustavsson_improving_2013}.
Extracting fidelity information can often be useful in diagnosing performance
and problems with a device in lieu of full characterization \cite{kelly_optimal_2014}.
Moreover, randomized benchmarking
has also been used to extract information
about the completely positive and unital parts of linear maps
\cite{kimmel_robust_2014}.

Here, using near-optimal data processing together with prior information,
we accelerate the data processing
used in benchmarking experiments, such that to achieve the accuracy demanded
of benchmarking protocols, we require orders of magnitude less experimental data.
We also extend results on the achievable estimation quality in the
presence of finite sampling \cite{epstein_investigating_2014}
and prior information, then show that our accelerated methods are nearly optimal.
Our data processing methods also provide estimates of their own performance,
such that our approach thus enables randomized benchmarking to be used where data
collection costs make existing benchmarking protocols impractical.
Thus, our work complements recent results on the robustness of randomized
benchmarking \cite{wallman_randomized_2014} to provide an
experimentally useful tool.

Randomized benchmarking has been recently used to adaptively calibrate
control designed by optimal control theory methods such as GRAPE
\cite{khaneja_optimal_2005}, allowing for differences between the control
model and the actual system to be adjusted for in experimental practice
\cite{egger_adaptive_2014}. These methods are applied in a control design
and calibration step, however, and do not allow for control for
to be recalibrated \emph{dynamically}. Whereas randomized benchmarking is performed at the
inner-loop of current control calibration algorithms \cite{kelly_optimal_2014},
any data collection overhead in benchmarking becomes a very significant
cost to control calibration as a whole.
Thus, by reducing the data requirements using
both better fitting methods and strong prior information, we can enable
new applications, such as extending control calibration to an online
context.

Here, we show that by using prior information together with
the sequential Monte Carlo (SMC) parameter estimation algorithm, we can obtain very
accurate estimates even in the limit of one bit of data per sequence length,
using instead a variety of sequence lengths to probe the
performance of our gate set.
We also show that for gates with fidelities near unity,
increasing the length of benchmarking sequences offers little compared to repeating experiments at already optimal sequence lengths.  The SMC algorithm is based on Bayesian methods,
which have been used successfully in a variety of quantum information processing
tasks \cite{schirmer_quantum_2010,schirmer_two-qubit_2009,huszar_adaptive_2012,sergeevich_characterization_2011,ferrie_how_2013,shulman_suppressing_2014,combes_-situ_2014}.
SMC has recently been used in
quantum information to learn states \cite{huszar_adaptive_2012} and Hamiltonians
\cite{granade_robust_2012,stenberg_efficient_2014}, and to provide robust error bounds
on inferred parameters \cite{ferrie_high_2014}. The primary cost incurred
by the SMC algorithm is that the data must be simulated repeatedly;
though this can be mitigated by using quantum resources
\cite{wiebe_hamiltonian_2014,wiebe_quantum_2014,wiebe_quantum_2014-2}.  Here we show that
since the symmetries afforded by random benchmarking experiments
can be used to simulate datasets with costs that are constant with
respect to the dimension of the Hilbert space of interest
\cite{magesan_characterizing_2012}, SMC can be implemented with little overhead.
Thus, randomized benchmarking mitigates the primary disadvantage
of SMC by removing the need to simulate the quantum system.

Moreover, the method of hyperparameters \cite{granade_robust_2012}
generalizes our approach to allow gate
fidelities to be non-trivial functions of some other parameter of interest,
such that the underlying parameter is learned directly.
This approach is especially relevant if, for example,
the effect of the unknown hyperparameter depends on an
experimental choice, such that distinct
benchmarking experiments can be used in concert in a straightforward way.

Our work proceeds first by defining the benchmarking model that we use,
then showing bounds on the estimation of the parameters of this model
using the Cramer-Rao bound. We then apply sequential Monte Carlo to
the benchmarking model and compare to the performance of traditional
methods, and to the optimal performance achievable with prior information,
showing that our method offers distinct advantages,
and is nearly optimal.

\section{Interpretation of Likelihood as Marginalization}
\label{sec:likelihood-interpretation}

Magesan \emph{et al} \cite{magesan_characterizing_2012} showed that the
average fidelity $F_g$ taken over all randomized benchmarking sequences of a
given length can be expressed in terms of the \emph{survival probability}
\begin{equation}
  \label{eq:survival-defn}
  \Pr(\text{survival} | \psi, \vec{i}_m) = \Tr[E_\psi \hhat{S}_{\vec{i}_m}(\rho_\psi)],
\end{equation}
where $E_\psi$ is a measurement effect corresponding to a fiducial state
$\rho_\psi$, and where
$\hhat{S}_{\vec{i}_m} = \hhat{S}_{i_m} \circ \cdots \circ \hhat{S}_{i_1}$
is the superoperator representing the sequence $\vec{i}_m$.
In particular, the expectation value of this survival probability over all sequences
of a given length $m$ was shown to produce the uniform-average fidelity
\begin{equation}
 F_g(m, \psi) = \expect_{\vec{i}_m | m}[\Pr(\text{survival} | \psi, \vec{i}_m)]
              = \Pr(\text{survival} | \psi, m).
\end{equation}
We may thus interpret the fidelity averaged over a unitary design as a
probability of survival in an experiment in which we do not know the sequence
being performed. As discussed in detail in \app{sampling-var},
if sequences are fairly drawn from the 2-design independently
of all other experimental choices, then this is a valid assumption, such that
the marginalized survival probability can be taken as the likelihood for our
randomized benchmarking model. Note that in the remainder of the paper, we
will let $\psi$ be fixed, and will drop the notation conditioning on this
assumption.

\begin{figure*}
    \centering
    \includegraphics[width=0.85\textwidth]{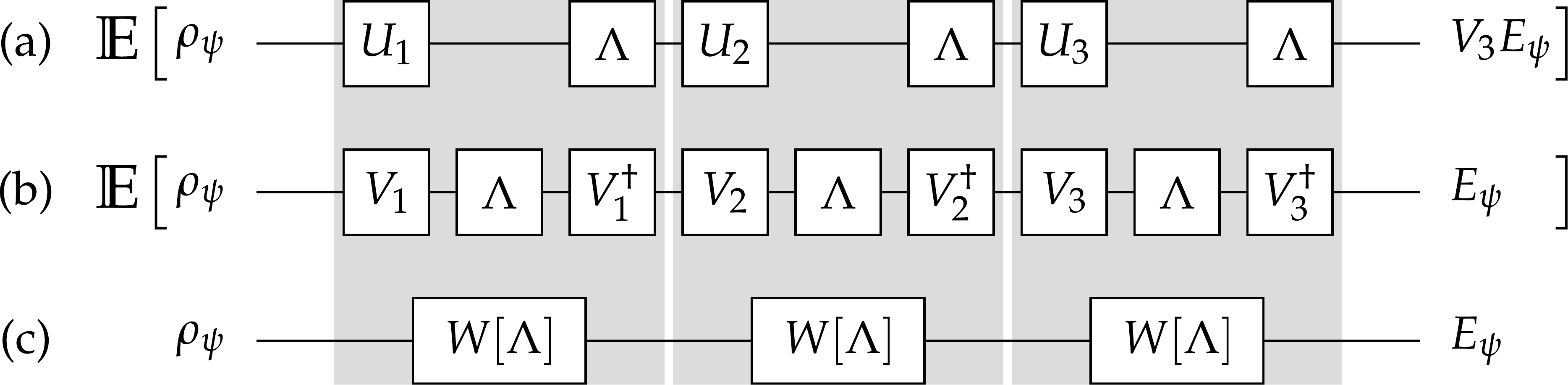}
    \caption{
        \label{fig:zeroth-deriv}
        Sketch of Magesan et al derivation of the zeroth-order model \cite{magesan_characterizing_2012}.
        (a) Sequence of length $m=3$ Clifford operations.
        (b) Change of variables to $V_i$, factoring out previous gates $U_{i-1}$,
            and with the base case $U_1 = V_1$.
            The $V$ gates then form a 2-design.
        (c) Expectation value over random gates in (a) and (b), giving
            the twirling superchannel $W$ acting on $\Lambda$.
    }
\end{figure*}

Using the expansion of the marginalized survival $F_g(m)$ given by Magesan
et al \cite{magesan_characterizing_2012}, we can rewrite the likelihood in a
way that explicitly depends on the parameters
of interest, and that no longer requires simulating the quantum dynamics
of the system. Thus, we can use Bayesian methods without simulating the
system under study.
In particular, we consider the zeroth-order model
\begin{equation}
  \label{eq:survival-zeroth}
  F_g(m) = A_0 p^m + B_0
\end{equation}
for parameters $A_0$, $B_0$ and $p$ given by
\begin{subequations}
\label{eq:zeroth-order-definitions}
\begin{align}
    A_0 & := \Tr\left[E_\psi \Lambda\left(\rho_\psi - \frac{\ident}{d}\right)\right] \\
    B_0 & := \Tr\left[E_\psi \Lambda\left(\frac{\ident}{d}\right)\right] \\
    p   & := (d F_\ave - 1) / (d - 1),
\end{align}
\end{subequations}
and where $F_\ave$ is the fidelity of the average channel
$\Lambda = \expect_{i,j}[\Lambda_{i,j}]$, taken over time steps $i$ and
elements of the gate set $j$. By these definitions, for ideal preparation,
evolution and measurement, $A_0 = 1 - \frac{1}{d}$ and $B_0 = \frac{1}{d}$.
Since we will often use the example of a qubit, we thus have that the ideal
$A_0 = B_0 = 1/2$. A sketch of the derivation of this model is given in
\fig{zeroth-deriv}.
The interpretation of first- and higher-order
models follows in a similar manner. Since we use the zeroth-order model as
an example in this work, we will drop the subscript-$0$ for brevity.

Because the fidelity of a channel is invariant under Clifford twirling,
the parameter $p$ represents the strength of the depolarizing channel
of fidelity $F_\ave$ produced by twirling the average channel $\Lambda$,
and can be used to recover $F_\ave$.
Similarly, in the interleaved protocol \cite{magesan_efficient_2012}, we consider two
probabilities, $p_{\text{ref}}$ and $p_{\bar{\mathcal{C}}}$,
respectively representing the sequences with $m$ random Clifford gates
multiplied together, or interleaved with some gate $\mathcal{C}$ under study.
From these probabilities, we can extract the referenced probability of
gate error $\tilde{p} \mathrel{:=} p_\rref / p_{\bar{\mathcal{C}}}$. Each of $p_\rref$
and $p_{\bar{\mathcal{C}}}$ is traditionally extracted
from a fit to the zeroth- or first-order model \footnote{
  We note that following the central limit theorem, the estimators
  $\hat{p}_\text{ref}$ and $\hat{p}_{\bar{\mathcal{C}}}$ will be approximately
  normally distributed about the true values of each parameter. Thus, estimating
  $\tilde{p}$ from $\hat{p}_\rref / \hat{p}_{\bar{\mathcal{C}}}$ results in
  an estimator that is Cauchy-distributed and therefore has no defined mean or
  variance. Estimates of the bias or error for this procedure therefore cannot be robustly
  provided by considering the sample standard deviations reported by least-squares
  fitting software.
}.

\section{Achievable Accuracy}
\label{sec:analytic-fisher}

We now consider only the interleaved model since it is more general.
For brevity, we represent the model by
a vector $\vec{x} = (\tilde{p}, p_\text{ref}, A, B)$, so that
the likelihood function for the interleaved model is
\begin{equation}
  \label{eq:interleaved-like}
  \Pr(1 | \vec{x}; m, \text{mode}) = \begin{cases}
      A p_{\text{ref}}^m + B & \text{mode is reference} \\
      A (p_{\text{ref}} \tilde{p})^m + B & \text{mode is interleaved}
  \end{cases}
\end{equation}
where we have labeled the survival event by ``1'' to more easily allow
for using binomial distributions to consider sums over multiple measurements
of the same sequence length.

Having defined our model, it is critical to account for the accuracy
with which we can estimate the parameters using finite data records.
Here, we extend the results of Epstein \emph{et al} \cite{epstein_investigating_2014} by
explicitly calculating the Fisher information of $\Pr(1|\vec{x})$.
We can find a bound on the achievable estimation error in this model
by appealing to the Cram\'er-Rao Bound \cite{cover_elements_2006},
which states that the Fisher information matrix $\matr{I}(\vec{x})$ bounds the
error matrix $\matr{E}(\vec{x})$ of \emph{any} unbiased estimator $\hat{\vec{x}}$ by the inequality
\begin{equation}\label{eq:crb}
\vec{E}(\vec{x}):=\mathbb E_{D|\vec{x}}[(\hat{\vec{x}}(D) - \vec{x})(\hat{\vec{x}}(D) - \vec{x})^{\T}]\geq \matr{I}(\vec{x})^{-1}.
\end{equation}
If the Fisher information matrix is singular, as is the case here
when all of the measured sequences are of the same length, the inverse is taken to be the Moore-Penrose
pseudo-inverse.
With at least four different sequence lengths, however,
we can break the degeneracy. Since this number depends
on the dimension of the model and not the underlying Hilbert
space, only four measurements are required to break the degeneracy,
even for systems of higher dimension than qubits.

It is often the case that we are only interested in $\tilde p$, the survival probability, and hence the gate fidelity, of a particular gate \cite{ryan_randomized_2009}.
In this case, we can bound the error of only that parameter by looking at a single element of the error and Fisher information matrix as
\begin{equation}
\vec{E}(\vec{x})_{\tilde p,\tilde p} \geq  1/\vec{I}(\vec{x})_{\tilde p,\tilde p}.
\end{equation}

\begin{figure*}[t!]
    \centering
    \includegraphics[width=0.3\textwidth]{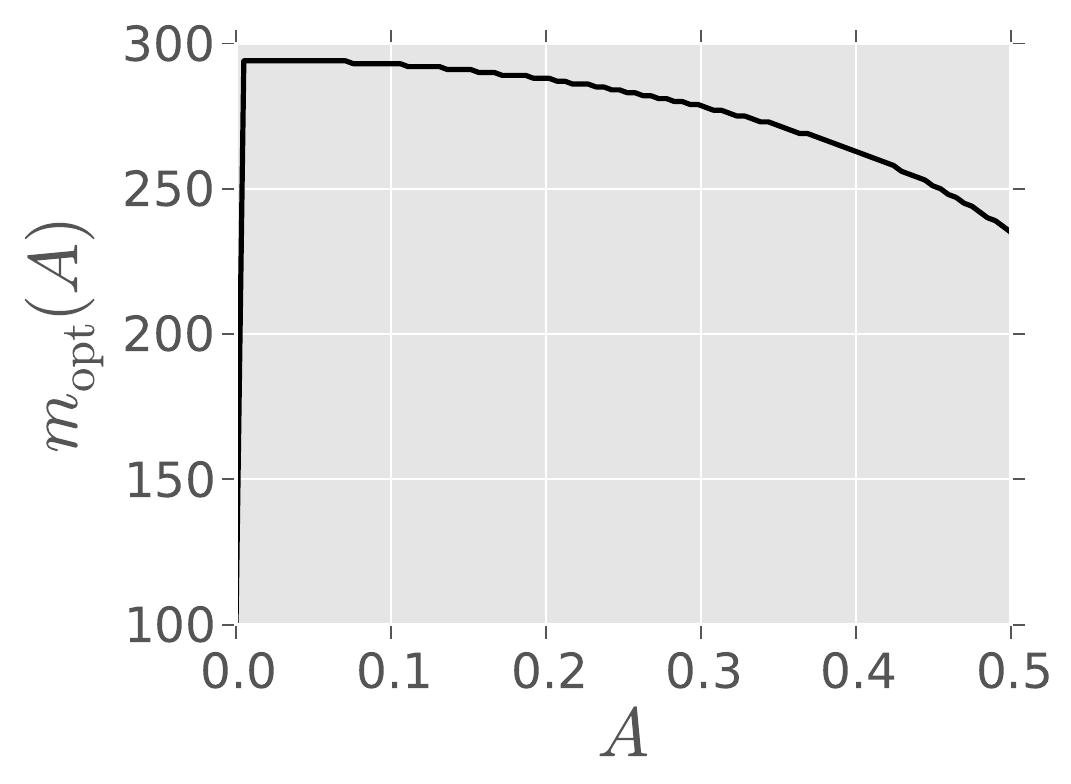}
    \hspace{1em}
    \includegraphics[width=0.3\textwidth]{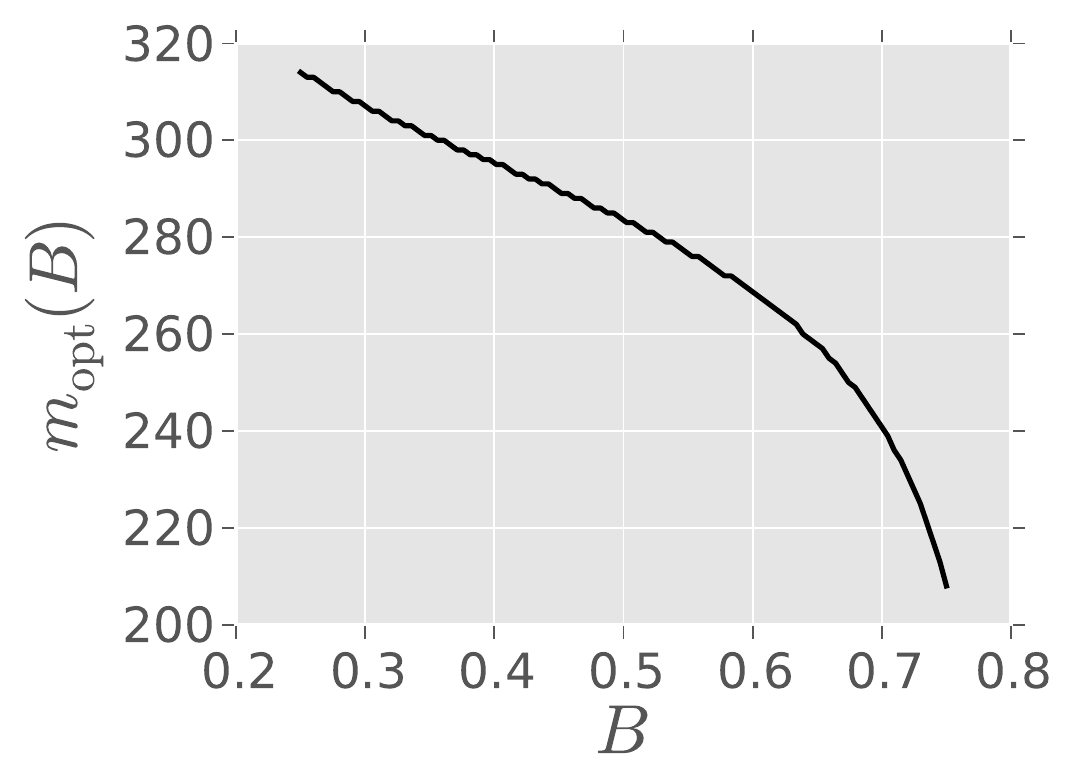}

    \includegraphics[width=0.5\textwidth]{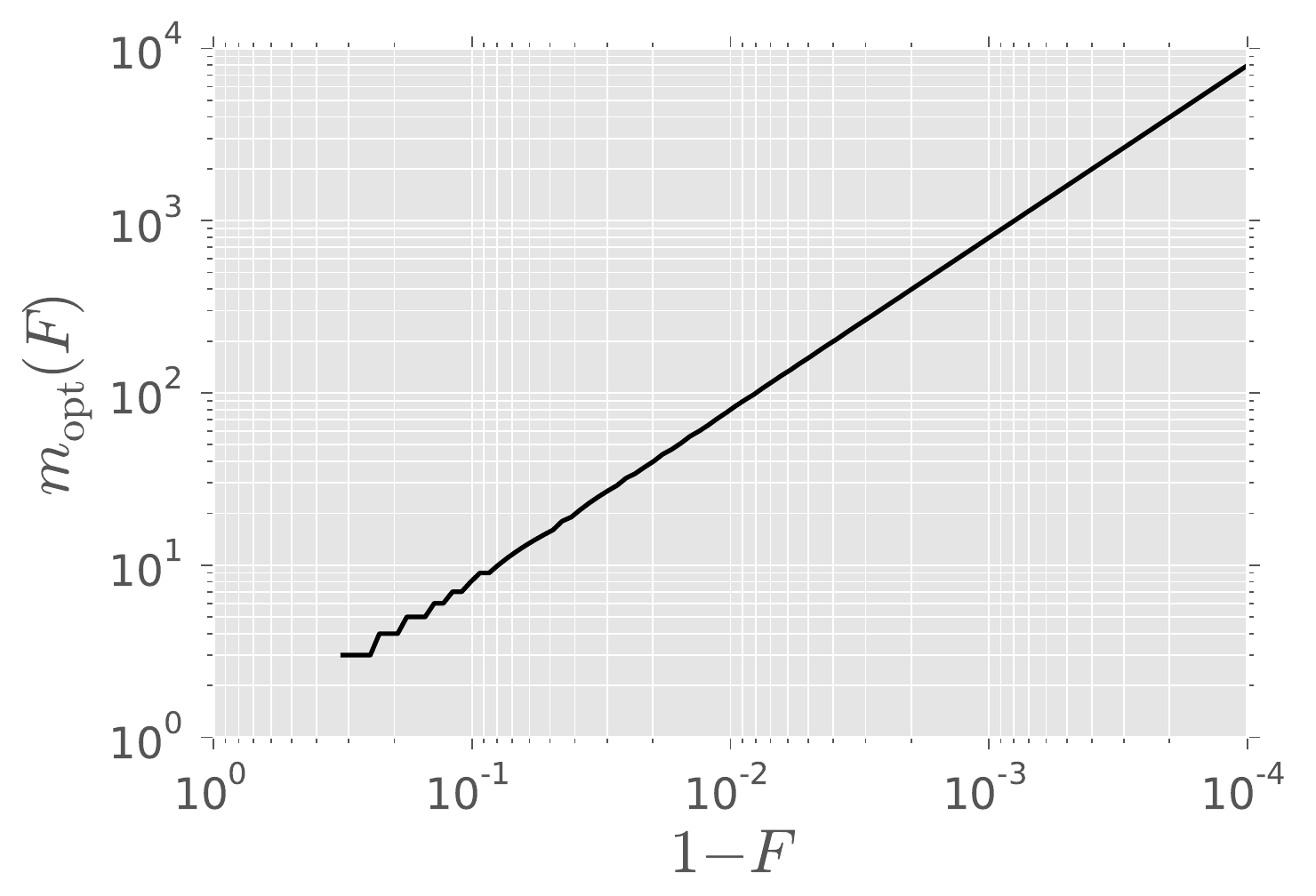}
    \caption{
        \label{fig:best-m}
        Optimal value of $m$ as a function of the scale $A$ and offset $B$ parameters,
        with $\tilde{p} = 0.9988$ and $p_\rref = 0.9978$, based on the example of \cite{barends_logic_2014}.
        On the top left, $B = 0.5$.  On the top right, $A$ = 0.25.
        Below, we take the limit as $d\to\infty$ of $\lceil m_{\text{opt}}\rceil$, assuming $\tilde{F} = F_\rref = F$.
    }
\end{figure*}

\begin{widetext}
To find the Fisher information for randomized benchmarking,
we derive the Fisher score $\vec{q}$ of this model, conditioned on $1$,
\begin{gather}
  \label{eq:fisher-score}
  \begin{aligned}
    \vec{q}(\vec{x} | 1; m, \text{mode})
          & = \nabla_{\vec{x}} \log \Pr(1 | \vec{x}; m, \text{mode}) \\
          & = \Pr(1 | \vec{x}; m, \text{mode})^{-1} \begin{cases}
              (
                0,\,
                A m p_{\text{ref}}^{m - 1},\,
                p_{\text{ref}}^m,\,
                1
              ) & \text{reference} \\
              (
                A m \tilde{p}^{m - 1} p_{\text{ref}}^m,\,
                A m \tilde{p}^m p_{\text{ref}}^{m - 1},\,
                p_{\text{ref}}^m,\,
                1
              ) & \text{interleaved}
          \end{cases},
  \end{aligned}
\end{gather}
where the similar expression for the outcome ``0'' follows immediately.
With this, we can calculate the Fisher information matrix
$\vec{I}(\vec{x}) \defeq \mathbb E_{D|\vec{x}}[\vec{q}(\vec{x} | D)\vec{q}(\vec{x} | D)^{\rm T}]$,
where $D$ labels the outcomes.
\end{widetext}

Fisher information analysis is one of the most powerful tools of statistical analysis since it bounds the performance of the continuous infinity of possible estimators we could choose.  However, given the difficulty of analytically computing the inverse of sums over Fisher information matrices of this form, we use numerical methods for its evaluation.
In particular,  QInfer \cite{granade_qinfer:_2012} performs this calculation automatically, given an implementation
of \eq{fisher-score}.


In experimentally relevant regimes, the task is to gain further accuracy when it is known \emph{a priori} that the fidelity is high.   To minimize the error in estimating $\tilde p$, we maximize the corresponding element of the Fisher information matrix.  
Note that, as is shown in \fig{best-m}, this optimum depends strongly on
the value of $A$ and $B$ when $\tilde p, p_\rref\approx 1$.  For ideal measurements and unital channels, as $d\to\infty$ we have $B\to 0$ and $A\to 1$ such that for large systems,
\begin{equation}
    \label{eq:best-m-large-d}
    \lim_{d\to\infty} m_{\text{opt}} \approx  \frac{1}{1 - \tilde{F} F_{\rref}}.
\end{equation}
As shown in \fig{best-m}, this can grow large for $|1 - F| \to 0$,
but even for fidelities near thresholds, such as $|1 - F| \approx 10^{-3}$
as considered by \cite{barends_logic_2014},
$m_{\text{opt}}$ remains manageable at about 500.


The above calculation is relevant in scenarios where the parameters not of
interest (that is, $A$ and $B$) are known fairly well and the gate fidelity is
already known to be near unity.  If we have prior information
that is not of this form, Bayesian analysis is better suited to the
task.

The Bayesian analogue of Fisher information analysis is a straightforward generalization.
We begin with a distribution $\pi(\vec{x})$, called a prior, over the parameters.
Ideally, this is a faithful encoding of the the experimenter's prior information, but the following analysis works equally well for \emph{any} distribution.  
In particular, given a prior distribution $\pi(\vec{x})$, the Bayesian information matrix $\matr{J}$ is then defined as
\cite{dauwels_computing_2005}
\begin{equation}
    \label{eq:bim-def}
    \matr{J} := \expect_{\vec{x}\sim\pi} [\matr{I}(\vec{x})].
\end{equation}
To calculate this we can perform a Monte Carlo integral over the prior by drawing samples $\vec{x}\sim\pi$ and evaluating $\matr{I}$ at each $\vec{x}$.

The Bayesian Cramer-Rao Bound (BCRB) then states that the error matrix
$\matr{E} := \expect_{\vec{x},D}[(\hat{\vec{x}}(D) - \vec{x})(\hat{\vec{x}}(D) - \vec{x})^{\T}]$
of any estimator $\hat{\vec{x}}$ satisfies
\begin{equation}
    \label{eq:bcrb}
    \matr{E} \ge \matr{J}^{-1}.
\end{equation}
The calculation of the BCRB is naturally included into the sequential Monte Carlo algorithm, 
such that our approach bounds its own performance based on the best experimental data
available.
Moreover, contrary to the Cramer-Rao bound in Eq.~\eqref{eq:crb}, it is known that the mean of the posterior distribution minimizes the error \cite{lehmann_theory_1998}. 
Thus, we need not seek the optimal estimator, as it naturally arises from
a representation of the posterior.

\section{Numerical Examples}
\label{sec:numerical-examples}

In the numerical examples we consider here, we choose $\pi$ to be a normal 
distribution with a mean vector $(\tilde{p}, p_\rref, A, B) = (0.95,0.95,0.3,0.5)$
and equal diagonal covariances given by a deviation of $\sigma= 0.01$.
The least-squares fit estimator is seeded with an initial guess drawn from
this prior, so as to fairly compare the estimators.
This distribution is intersected with the hard constraints implied by definitions of the parameters, which defines the support of the prior as
\begin{align}
\supp\pi = \{(A,B,p): &-1\leq A\leq 1, 0\leq B\leq 1, \\
&0\leq p \leq 1, 0\leq Ap+B\leq 1\}.\nonumber
\end{align}
This distribution
was chosen as the likelihood model is less degenerate given these constraints,
such that it is easier to reason about bounds for approximately unimodal
estimation strategies. Our choice of prior is not critical, however,
as we will show later that our algorithm recovers well from the case
in which we choose a ``bad'' prior.

\begin{figure}
  \centering
  \includegraphics[width=0.49\textwidth]{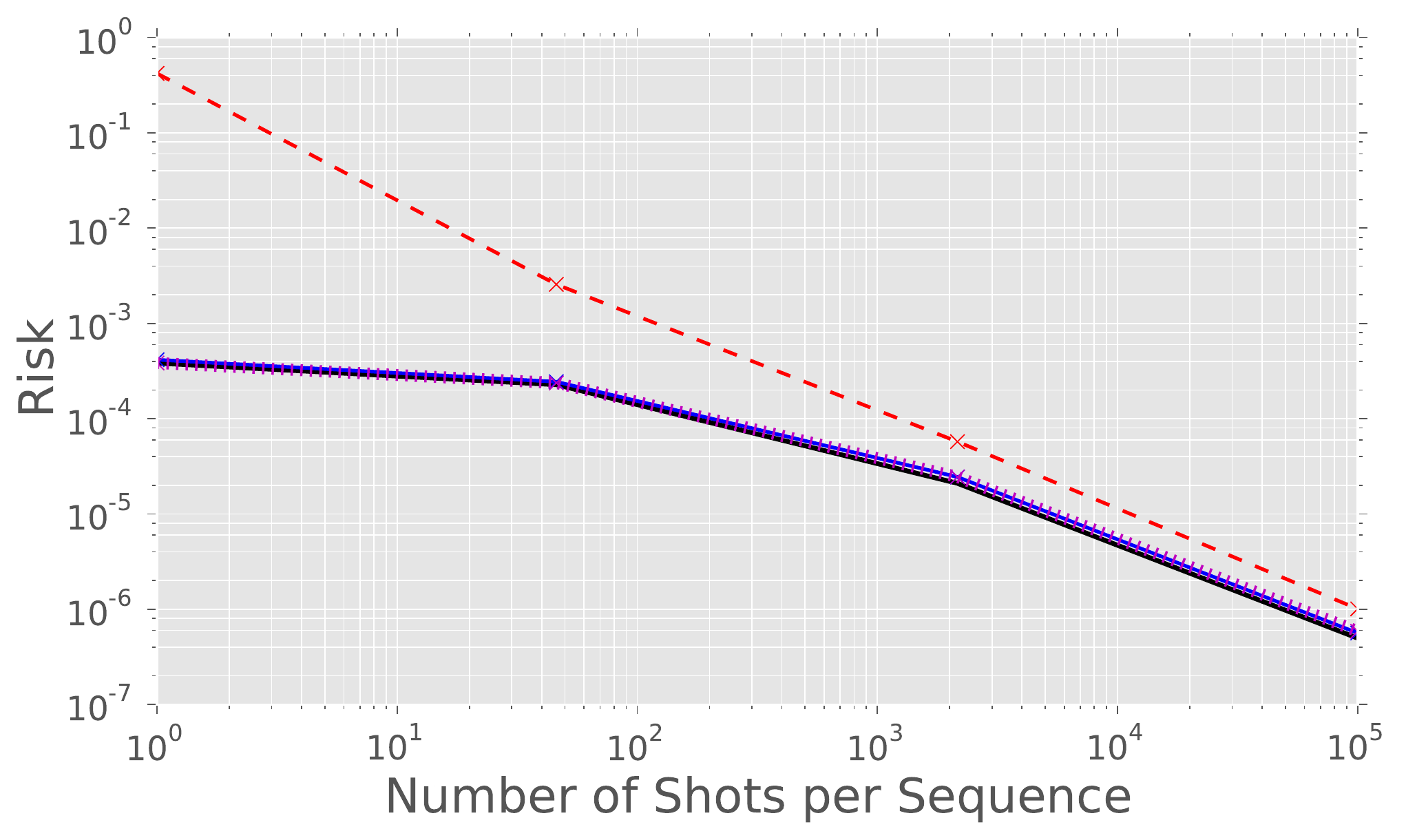}

  \includegraphics[width=0.49\textwidth]{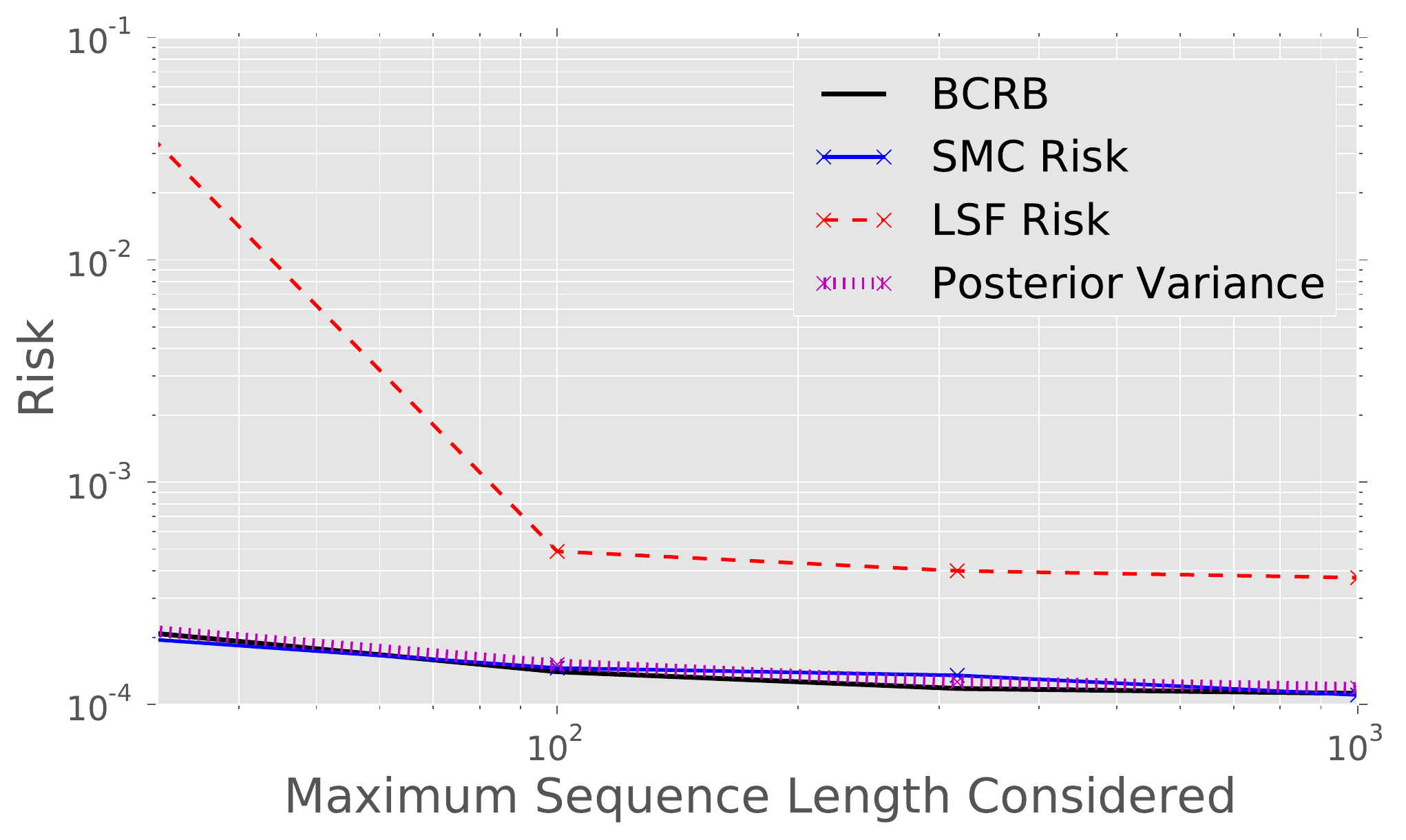}
  \caption{
    \label{fig:risk-comparison}
    Comparison of mean squared error achieved by sequential Monte Carlo (SMC) and least-squares
    fit (LSF) estimators of  $\vec{x} = (\tilde{p}, p_\rref, A, B)$, averaged over 100 trials and varied over
    the number of sequences per length $K$ (top) and the number of the maximum sequences $m_{\rm max}$ (bottom). The Bayesian Cramer-Rao Bound and
    posterior variance (an estimate of SMC's performance) are also shown.  On top, the simulated reference signal was taken with sequence lengths $\{1,2,\ldots,100\}$ and the interleaved signal was taken with $m \in \{1,2,\ldots,50\}$.  On the bottom, $K = 10^3$ samples were simulated per sequence length.  For each $m_{\rm max}$, $m\in\{1,11,21,\ldots,m_{\rm max}\}$ was chosen.
  }
\end{figure}

To demonstrate the Bayesian approach, we compare the standard least squares fit (LSF) performance
to the sequential Monte Carlo (SMC) algorithm \cite{granade_robust_2012},
which computes estimates by updating the probability of each of a finite
list of hypotheses according to Bayes' rule. In the case of randomized benchmarking,
this consists of computing \eqref{eq:interleaved-like} for each hypothesis
after each batch of measurements. We note that the cost of computing
\eqref{eq:interleaved-like} is independent of the dimension of the system,
such that randomized benchmarking explicitly avoids simulating quantum evolution
with classical resources.

There are essentially two experimental design choices an experimenter can make: the length of the sequence $m$, and the number of repetitions $K$.  In the first comparison, we fix the sequence length and vary $K$.  In particular, we take all sequence lengths up to 100 for the reference signal and 50 for the interleaved signal. For each
such $K$, we plot the mean squared error for the SMC and LSF estimators,
along with the posterior variance, which provides an online estimate of the performance of SMC, and the
Bayesian Cramer-Rao Bound. The results,
shown in \fig{risk-comparison}, demonstrate that SMC can be used to
obtain useful estimates of $\tilde{p}$ with a \emph{few orders of magnitude}
less data than is used by least-squares fitting. Moreover, this advantage
becomes more pronounced as the number of shots per sequence length
approaches one, such that SMC is especially useful in cases where data
collection is expensive. We note that this advantage reflects both the
performance of SMC itself, and the ability of SMC to take advantage of prior
information: for small amounts of data, the least-squares fit estimator
chooses estimates far from the initial guess drawn from the prior
distribution, while the SMC estimate instead refines the prior.
Moreover, SMC can accurately characterize its own performance
and can obtain significantly closer accuracy
to the ultimate bound given by the BCRB. 
These advantages are similar to other cases in which
SMC shows a large advantage over traditional fitting methods
for handling data that is far from deterministic
\cite{granade_robust_2012,ferrie_likelihood-free_2014}.

\begin{figure*}
    \centering
    \includegraphics[width=0.57\textwidth]{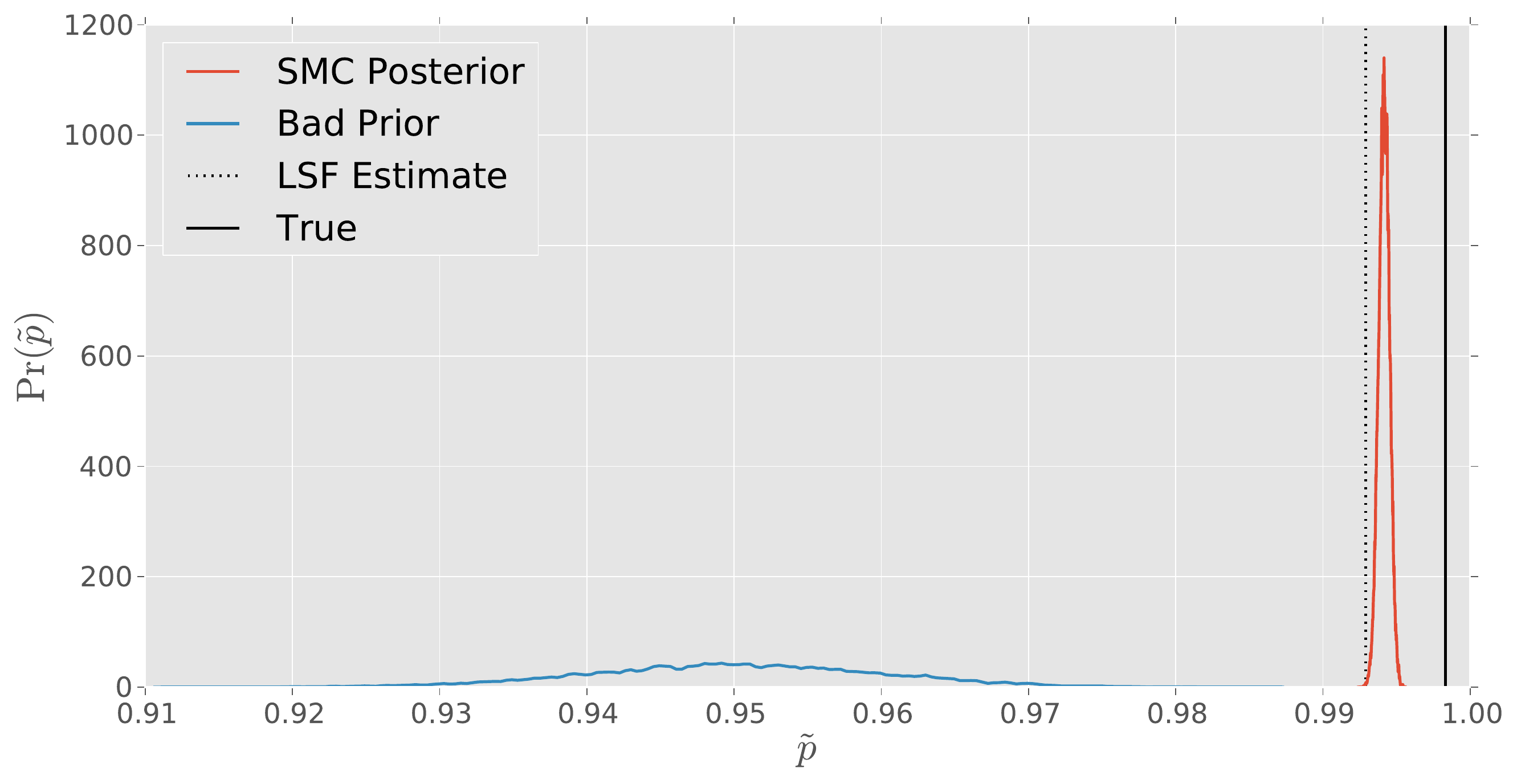}
    \includegraphics[width=0.42\textwidth]{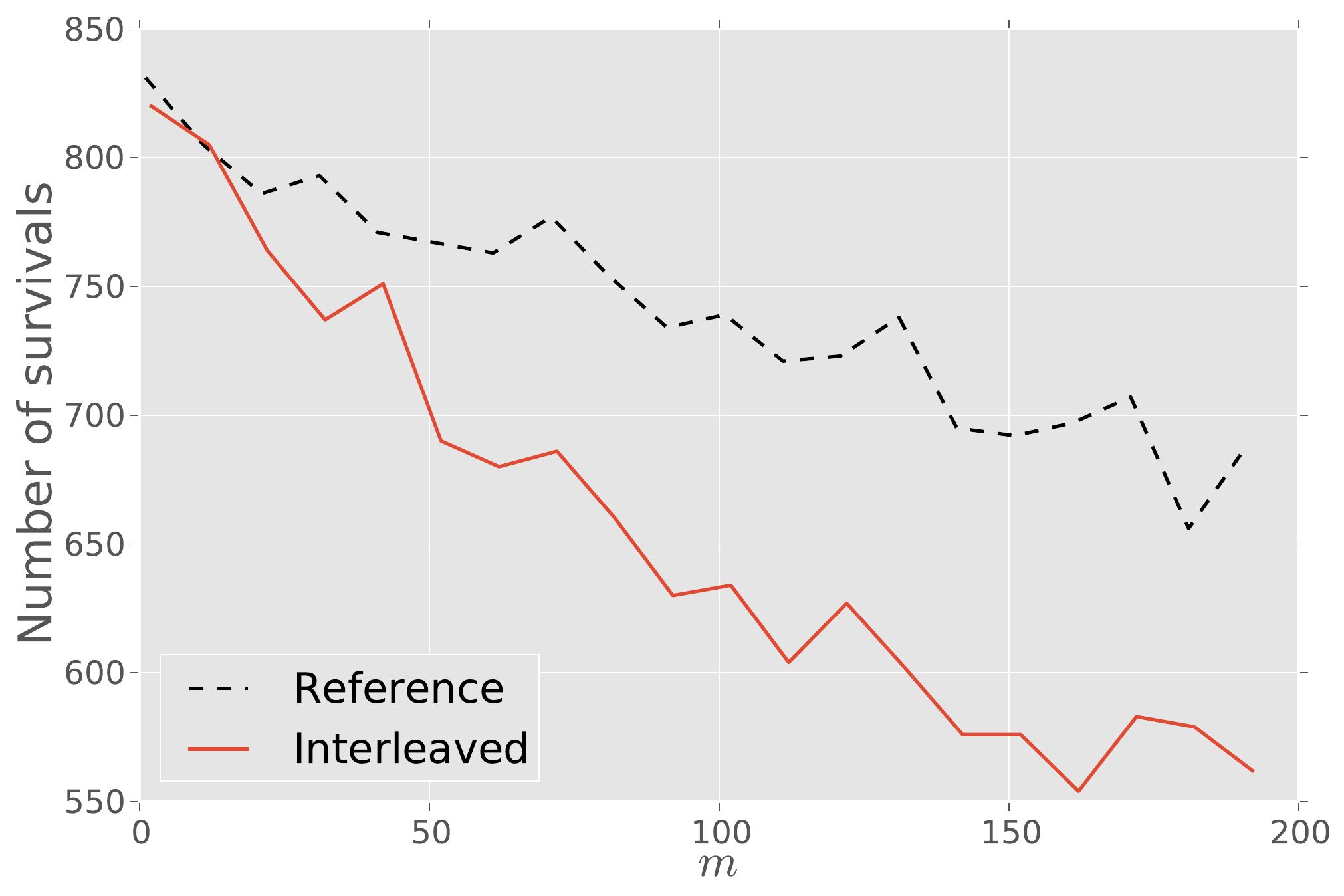}

    \caption{
        \label{fig:bad-prior}
        (Left) Comparison of prior distribution,
        SMC-approximated posterior, true value and LSF-estimate for $\tilde{p}$
        for a single run with $K = 1000$ shots at each of $m_{\text{ref}} \in
        \{1, 11, \dots, 191\}$ and $m_{C} \in \{2, 12, \dots, 192\}$.
        An intentionally inaccurate prior is used, such that the true value is
        approximately 6.9 standard deviations from the mean of the prior.
        As shown in \tab{physsim-results}, SMC does well by comparison to LSF,
        even with the poorly-chosen prior. \\
        (Right) Data gathered from simulation with physical-model gates.
    }
\end{figure*}

\begin{figure*}
    \centering
    \includegraphics[width=0.57\textwidth]{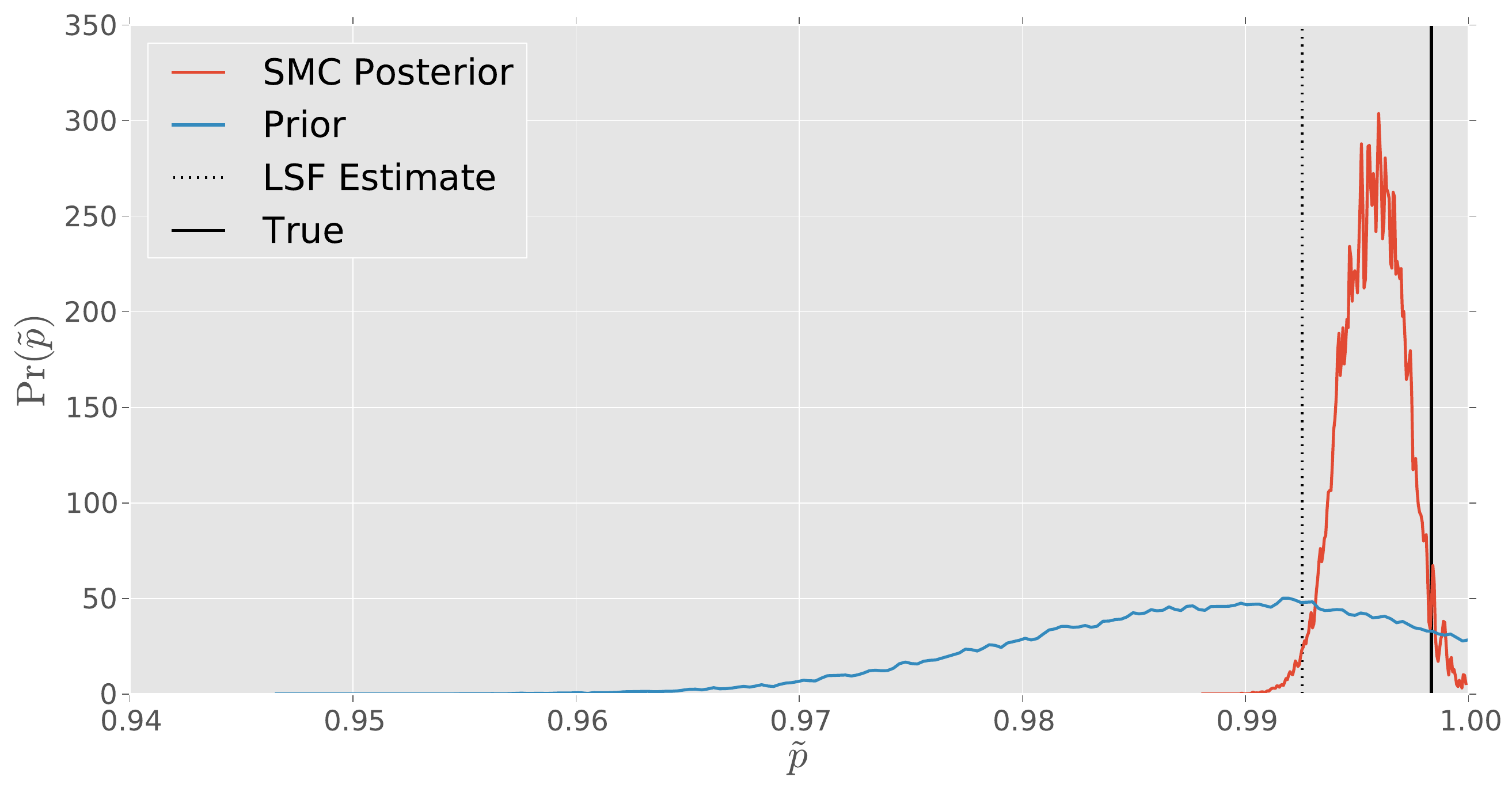}
    \includegraphics[width=0.42\textwidth]{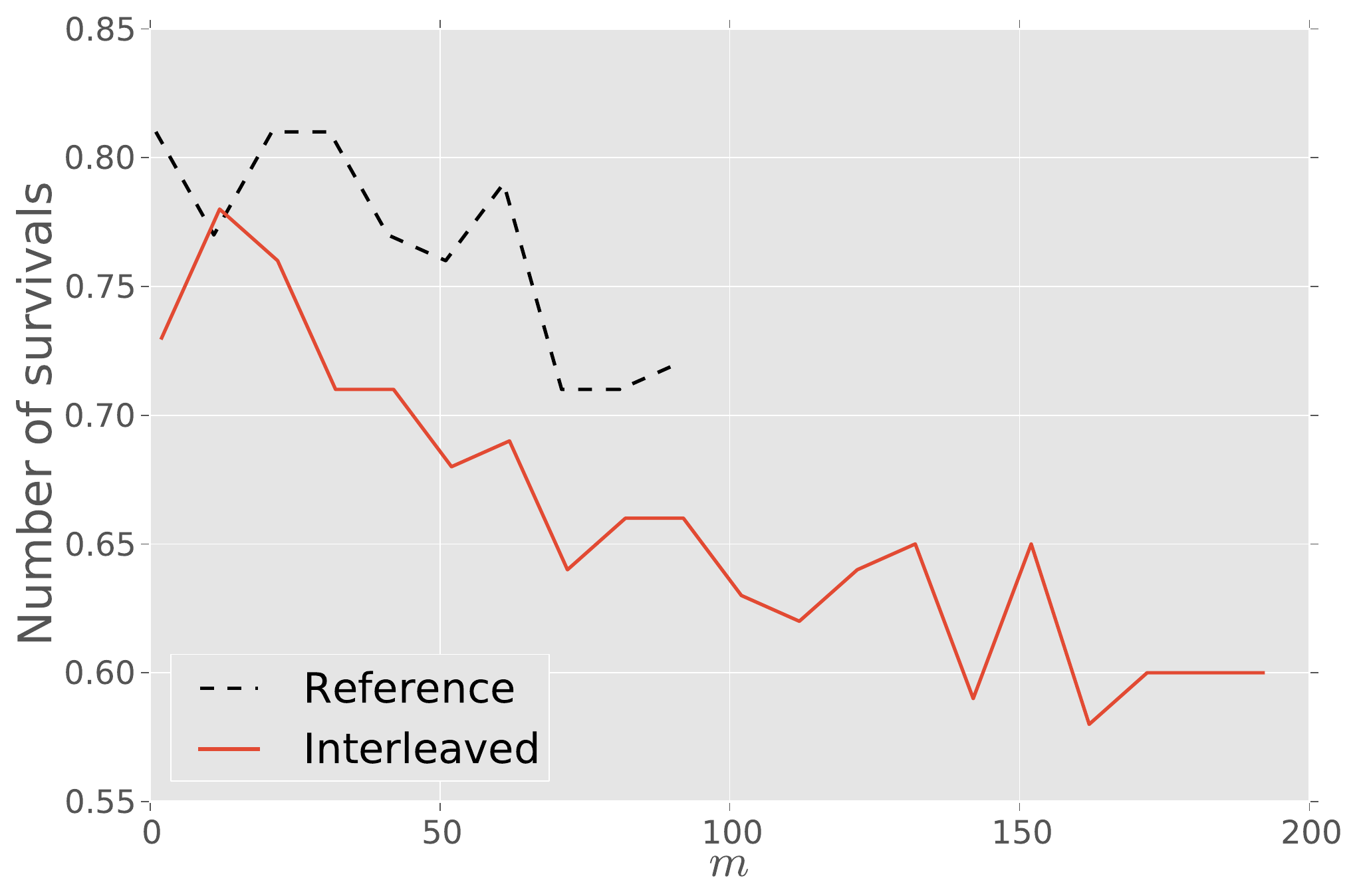}

    \caption{
        \label{fig:good-prior}
        (Left) Comparison of prior distribution,
        SMC-approximated posterior, true value and LSF-estimate for $\tilde{p}$
        for a single run with $K = 100$ shots at each of $m_{\text{ref}} \in
        \{1, 11, \dots, 91\}$ and $m_{C} \in \{2, 12, \dots, 192\}$. \\
        (Right) Data gathered from simulation with physical-model gates.
    }
\end{figure*}

In \fig{risk-comparison} (bottom), we show the performance of SMC and LSF when the sequence lengths $m$ vary and the number of shots $K$ per sequence length is fixed, demonstrating that SMC can improve upon LSF especially for very short sequences.  Moreover, we see the benefit from increasing the sequence length is \emph{minimal} compared to repeating experiments at a given sequence length near the optimum length found from the Cramer-Rao bound.

\section{Benchmarking with Simulated Gates}
\label{sec:phys-benchmarking}

Thus far in the analysis, we have used as a simulator the same zeroth-order model as is used to process and interpret the data. To demonstrate the utility of our approach in comparison with traditional LSF-based benchmarking,
we now simulate gates according to a cumulant expansion, with physically
realistic models. In particular, we use the superconducting model of
\cite{puzzuoli_tractable_2014} together with optimal control theory
\cite{khaneja_optimal_2005} to generate a set of gates implementing
the target unitaries $\{\ident, X, Y, Z, H, P\}$, where $H$ is the Hadamard
gate, and where $P = \ket{0}\!\!\bra{0} + \ii \ket{1}\!\!\bra{1}$ is the phase gate.
We then use the superoperators $\hhat{S}_{U}$ for implementing each target
unitary $U$ obtained from a cumulant simulation
\cite{kubo_generalized_1962,cappellaro_principles_2006} to 
sample from the likelihood function
\eqref{eq:survival-defn} \footnote{To ensure that the ideal action of each sequence
is the identity operation, we use Gottesman-Knill simulation \cite{aaronson_improved_2004}
as implemented by the QuaEC library \cite{granade_quaec:_2012}
to find the inverse of the first $(m-1)$ gates in each sequence,
and then set the $m^{\text{th}}$ gate to be the inverse. The
algorithm
for implementing the simulator
is described in the supplemental materials.}.

To process these samples, we then use the zeroth-order
likelihood function \eqref{eq:interleaved-like} both as a model for
sequential Monte Carlo and as a trial function for least-squares fitting. 
Since the actual implemented gates are known, we can compute the
true parameters for comparison.
In \tab{physsim-results}, we show the true parameters, the result obtained
using SMC, and the result obtained using least-squares fitting. 
The most important thing to note is that correct parameters are a distance 6.90 $\sigma$ from the prior (meaning the true parameters are outside of the 99.9999998\% credible ellipse).  This shows that even in the case when the prior information fails to accurately capture the uncertainty in the true model, SMC still does well,
providing evidence that our accelerated methods may also be \emph{robust},
even when used to measure the fidelities of sets of gates with errors
that are correlated between distinct gate types, or that include
non-trivial unitary components
 \footnote{Note that SMC did not act in a robust manner in all cases
observed, but in those cases where SMC did not
do well by comparison to LSF, the QInfer package was able to warn
by using the effective sample size criterion described
in \cite{granade_robust_2012}, such that the data processing
could then be repeated if necessary, or such that a more appropriate
prior could be chosen.}.
  We show this in more detail in
\fig{bad-prior}, comparing the posterior and prior distributions
over $\tilde{p}$ to the true and LSF-estimated values.

\begin{table*}
    \caption{
        \label{tab:physsim-results}
        Results of using SMC and least-squares fitting to estimate
        the fidelity of $U = X$, simulated using the superconducting qubit
        gate set. (Left) Bad prior from \fig{bad-prior},
        (right) accurate prior from \fig{good-prior}.
    }

      \begin{tabular}{l|cccc@{\hskip 2em}cccc}
          & \multicolumn{4}{c}{Bad Prior ($40 \times 10^3$ bits)} & \multicolumn{4}{c}{Good Prior ($3 \times 10^3$ bits)} \\
          \hline
          & $\tilde{p}$ & $p_{\text{ref}}$ & $A_0$ & $B_0$ & $\tilde{p}$ & $p_{\text{ref}}$ & $A_0$ & $B_0$ \\
          \hline
          {True}         & 0.9983 & 0.9957 & 0.3185 & 0.5012      & 0.9983 & 0.9957 & 0.3185 & 0.5012  \\
          {SMC Estimate} & 0.9942 & 0.9971 & 0.3023 & 0.5075      & 0.9957 & 0.9969 & 0.2973 & 0.5010  \\
          {LSF Estimate} & 0.9929 & 0.9974 & 0.3423 & 0.4827      & 0.9925 & 0.9986 & 0.5153 & 0.2782  \\
          \hline
          {SMC Error}    & 0.0042 & 0.0014 & 0.0161 & 0.0062      & 0.0026 & 0.0011 & 0.0212 & 0.0003 \\
          {LSF Error}    & 0.0054 & 0.0017 & 0.0239 & 0.0185      & 0.0058 & 0.0029 & 0.1968 & 0.2230
      \end{tabular}
\end{table*}

Finally, in \fig{good-prior}, we demonstrate the advantage of
our method in the presence of physical gates together with
a more reasonable prior, and using approximately 10-fold less data
than in \fig{bad-prior}. 
Taken with other evidence of the robustness of SMC methods \cite{wiebe_quantum_2014,ferrie_likelihood-free_2014},
these results thus show that our method is useful and provides advantages in data collection
costs in experimentally-reasonable conditions.

We also note that LSF provides an accurate estimate of $\tilde p$ for
the simulations with physical gates,
but it appears to be at the expense of providing poor estimates for $A$ and $B$.
Given that the errors in $\tilde{p}$ and those in $A$ and $B$ are not in general
uncorrelated, that LSF often provides such poor estimates of $A$ and $B$ makes the
estimates of $\tilde{p}$ derived from LSF difficult to trust.

In this work, we have discussed the fundamental limits of the randomized
benchmarking technique that are incurred due to small data sets, and have
shown an algorithm that reliably saturates this optimum. In doing so, we have
shown that by using sequential Monte Carlo, with a moderate tradeoff in
computational costs, one can obtain as much as two
orders of magnitude improvement in estimation accuracy, such that data collection
requirements are similarly reduced by as much as a hundred-fold. Given
the wide and expanding use of randomized benchmarking in experimental
practice, this then translates to a significant performance benefit both in
benchmarking, and in experimental protocols derived from benchmarking.


\begin{acknowledgments}
    The numerical methods used in this paper are implemented using the
    QInfer, QuaEC, QuTiP 2 and SciPy libraries for Python
    \cite{granade_qinfer:_2012,granade_quaec:_2012,johansson_qutip_2013,jones_scipy:_2001}.
    \ifthenelse{\boolean{JournalVersion}}{
        Literate source code for this work can be found at \footnote{The source code is available at \url{https://github.com/cgranade/accelerated-randomized-benchmarking}. To view the code online, visit \url{http://nbviewer.ipython.org/github/cgranade/accelerated-randomized-benchmarking/blob/master/src/model_testing.ipynb}.}.
    }{}
    We thank Holger Haas for simulating the gate sets used above,
    and for discussions. We thank Nathan Wiebe,
    Ben Criger, Ian Hincks, Josh Combes, Joel Wallman, Joe Emerson and Yuval Sanders for useful discussions and
    insights. 

    CG and DGC acknowledge funding from Industry Canada, CERC, NSERC
    and the Province of Ontario.
    CF acknowledges funding from NSF Grant No. PHY-1212445 and NSERC.
\end{acknowledgments}

\bibliographystyle{apsrev}
\bibliography{benchmarking}


\appendix
\onecolumngrid

\section{Sampling Variance and Derivation of Marginalized Likelihood}
\label{app:sampling-var}

\newcommand{\Cli}{\mathcal{C}}
\newcommand{\sop}{\hhat}
\newcommand{\len}{\ensuremath{\operatorname{len}}}
\newcommand{\var}{\ensuremath{\mathbb{V}}}

In this derivation, we will focus on the zeroth-order model of 
Magesan et al \cite{magesan_characterizing_2012}, which gives that the average
fidelity $F_g(m)$ over all sequences of length $m$ is given by
\begin{equation}
    F_g(m) = A_0 p^m + B_0
\end{equation}
for constants $A_0$ and $B_0$ describing the state preparation and measurement
(SPAM) errors, and where $1 - p$ is the depolarizing strength
of $W[\expect_{C\sim\Cli_n}\sop{S}_{C}]$.

We are interested in the single-shot limit, where each measurement consists of
first selecting a sequence, then measuring once the survival probability for
that sequence. Since this protocol makes no use of the sequence other than its
length, we can describe the protocol by marginalizing over the choice of
sequence, giving a probability distribution of the form
$\Pr(\text{survival} | m)$, where $m$ is a sequence length.

To derive this, we first pick a length $m$, and then consider the choice
of sequence $\vec{i}$ out of all length-$m$ sequences to be a random variate.
Thus, there exists probabilities
\begin{equation}
    p_{m, \vec{i}} \defeq \Pr(\text{survival} | \vec{i}, m) = \Tr(E_\psi \sop{S}_{\vec{i}}[\rho_\psi])
\end{equation}
for each individual sequence
that we could have chosen, such that marginalizing over results in
\begin{equation}
    \Pr(\text{survival} | m) = \expect_{\vec{i}}[\Pr(\text{survival} | \vec{i}, m)].
\end{equation}
If each sequence is drawn with uniform probability, then
\begin{equation}
    \Pr(\text{survival} | m) = \frac{1}{|\Cli_n|^m}\sum_{\vec{i}~\text{s.t.}~\len\vec{i} = m} p_{m, \vec{i}}.
\end{equation}
We recognize this as being the average sequence fidelity $F_g(m)$ modeled by
Magesan,
\begin{equation}
    \label{eq:arb-zeroth-likelihood}
    \Pr(\text{survival} | m) = F_g(m) = A_0 p^m + B_0.
\end{equation}

To interpret $F_g(m)$ as a likelihood directly, note that we had to consider
the Bernoulli trial (single-shot) limit; had we instead taken $K$ distinct
sequences and measured each $N > 1$ times, we would have arrived at a quite
different quantity
\begin{equation}
    \hat{F}_g(m) = \sum_{k = 1}^{K} \hat{F}(m, \vec{i}_k),
\end{equation}
where $\hat{F}(m, \vec{i}_k)$ is the estimate of the sequence fidelity for the
\emph{particular} sequence $\vec{i}_k$.

The difference is made clear by considering an example with fixed sequence length
$m$, and the variance for a datum $d \sim \Pr(\text{survival} | m)$ (labeling
``survival'' as 1 and the complementary event as 0),
\begin{equation}
    \var_d[d | m] = \var_{\vec{i}}[\expect_d[d | \vec{i}, m]] +
                    \expect_{\vec{i}}[\var_d[d | \vec{i}, m]].
\end{equation}
The second term corresponds to the mean variance over each fixed sequence
$\vec{i}_m$, and governs how well we can estimate each $F(m, \vec{i})$
individually. The first term, however, is more interesting, in that it measures
the variance \emph{over sequences} of the per-sequence survival probability
$p_{m,\vec{i}} = \expect_d[d | \vec{i}, m]$. By the argument of Wallman and Flammia
\cite{wallman_randomized_2014}, this is small when the fidelity being estimated is close to 1;
that is, when the gates being benchmarked are very good. For gates that
are farther from the ideal Clifford operators, however, or for applications
such as tomography via benchmarking \cite{kimmel_robust_2014},
this term is not negligible, mandating that many different sequences must be
taken for $\hat{F}_g(m)$ to be a useful estimate of $F_g(m)$.

By demanding that each individual shot be drawn from an independently chosen
sequence, our approach avoids this and samples from $d | m$ directly. In this
way, we see a similar effect as in \cite{ferrie_how_2013}. In particular, it is
not advantageous to concentrate one's sampling on one point,
but to spread samples out and gain
experimental variety. Here, the one shot per sequence limit plays the role of the
one sample per time-point limit in the earlier discussion.

\end{document}